\documentstyle[aps,epsf,multicol]{revtex}
\def\x{{\bf x}}
\def\r{{\bf r}}
\def\B{{\bf B}}
\def\q{{\bf q}}

\def\BRA{\left\langle}
\def\KET{\right\rangle}
\def\LK{\left(}
\def\RK{\right)}

\def\be{\begin{eqnarray}}
\def\ee{\end{eqnarray}}
\begin{document}
\draft \title{Magnetotunneling spectroscopy of mesoscopic
  correlations\\ in two-dimensional electron systems}

\author{Julia S. Meyer, Alexander Altland,
  and Martin Janssen} \address{Theoretische Physik III,
  Ruhr-Universit\"at Bochum, 44780 Bochum, Germany} \date{\today}
\maketitle
\begin{abstract}
  An approach to experimentally exploring electronic correlation
  functions in mesoscopic regimes is proposed. The idea is to monitor
  the mesoscopic fluctuations of a tunneling current flowing between
  the two layers of a semiconductor double-quantum-well structure.
  From the dependence of these fluctuations on external parameters,
  such as in-plane or perpendicular magnetic fields, external bias
  voltages, etc., the temporal and spatial dependence of various
  prominent correlation functions of mesoscopic physics can be
  determined. Due to the absence of spatially localized external
  probes, the method provides a way to explore the interplay of
  interaction and localization effects in two-dimensional systems
  within a relatively unperturbed environment.  We describe the
  theoretical background of the approach and quantitatively discuss
  the behavior of the current fluctuations in diffusive and ergodic
  regimes. The influence of both various interaction mechanisms and
  localization effects on the current is discussed. Finally a proposal
  is made on how, at least in principle, the method may be used to
  experimentally determine the relevant critical exponents of
  localization-delocalization transitions.
\end{abstract}
\pacs{PACS numbers:
73.21.-b, 
73.23.Hk, 
73.40.Gk, 
73.50.-h 
}

\begin{multicols}{2}
\narrowtext

\section{Introduction}\label{sec-int}
At low temperatures, disordered or chaotic electronic systems are
strongly affected by mechanisms of quantum interference. These
interference effects manifest themselves in anomalously strong
fluctuations of both thermodynamic and transport observables and in
the spatial localization of quantum mechanical wave
functions~\cite{reviews-meso}. They find their common origin in an interplay of the classical nonintegrability of the charge carrier
dynamics and the wave nature of quantum mechanical propagation.  To
quantitatively characterize
this `mesoscopic' behavior, one commonly employs correlation
functions of the type 
\begin{eqnarray*}
&&F(\x_1,\x_2,\x_3,\x_4;\omega,\Delta\alpha)
\equiv\\
&&\qquad \equiv\langle G^-(\x_1,\x_2;\epsilon,\alpha) G^+ (\x_3,\x_4;\epsilon+
\omega, \alpha+ \Delta\alpha) \rangle\,,
\end{eqnarray*}
where $G^\pm$ is the
retarded/advanced single-particle Green function.  Here $\langle \dots
\rangle$ stands for some kind of averaging (e.g.,  averaging over
realizations of disorder) and the parameter $\alpha$ symbolically
represents an optional dependence of the Green function on external
control parameters (magnetic fields, gate voltages or others).
Correlation functions of this type appear as the `most microscopic'
building block in the analysis of the majority of fluctuating
mesoscopic observables. As a consequence of constructive quantum
interference these objects become long ranged whenever the spatial
arguments $\x_1,\dots,\x_4$ are pairwise close (on scales of $l_{\rm
  min}$, the range of the averaged Green functions $\langle G^\pm\rangle$).  Specifically,
\begin{itemize}
\item For $\x \equiv \x_1\approx \x_2$ and $\x'\equiv \x_3 \approx
  \x_4$, $F^{[d]}(\x,\x') \equiv F(\x,\x,\x',\x')$ describes
the fluctuations of the (local) density of states (DoS), and, thus, the
thermodynamic fluctuations and parametric correlations.
\item For $\Delta \alpha=0$, $F^{[D]}(\x,\x') \equiv
  F(\x,\x',\x',\x)$ describes the total probability of propagation
  from $\x$ to $\x'$. This is the generalized `diffuson', a quantity of
  key relevance in the context  of mesoscopic transport.
\item Finally, in a system with unbroken time reversal symmetry,
  $F^{[C]}(\x,\x') \equiv F(\x,\x',\x,\x')$, the cooperon, becomes
  long ranged, too.
\end{itemize}

The dependence of the correlators $F^{[d;D;C]}$ on the long-ranged
distance $r\equiv|\x-\x'|$, the energy difference $\omega$, and
$\Delta\alpha$ fundamentally characterizes a multitude of mesoscopic
phenomena~\cite{reviews-meso}.  For this reason, many {\em
  theoretical} investigations in mesoscopic physics concentrate on an
analysis of these objects.  Experimentally, however, it has proven
difficult to access the correlation functions $F^{[d;D;C]}$ directly:
Ideally, one would like to {\em continuously} measure the dependence
of the correlators $F^{[d;D;C]}$ over a range of at least the
parameters $r$ and $\omega$.  Irritatingly, this cannot be achieved
within experimental setups based on a standard device-contact-electron
system architecture. In fact, the mere presence of local contacts
introduces an entire spectrum of difficulties obstructing the
continuous experimental spectroscopy of transport and spectral
correlation functions: First, the {\em fixed} attachment of local
current/voltage electrodes prevents one from continuously monitoring
the scale ($r$) dependence of transport correlation functions.  This
problem does not exist in measurements based on local tunneling
tips~\cite{STM}.  In those, however, the electronic state of the
tunneling device as well as its coupling to the electron system have
to be precisely known to draw quantitatively reliable conclusions on
the nature of the bulk electronic correlations of the latter. In
particular, the distance between the device and the electron system
has to be kept constant with atomic precision. These conditions can
hardly be met under realistic conditions.  Second, both local contacts
and tunneling tips tend to disturb the electron system under
investigation. In interacting systems, they lead to various
manifestations of the orthogonality catastrophe. As a consequence,
much of the measured current/voltage characteristics
describes the process of local accommodation of charge carriers at the
interface, rather than the electronic correlations of the bulk
system.   Third, a division between system and
contacts of a mesoscopic conductor is, to a large extent, arbitrary.
Quantum interference phenomena in mesoscopic systems tend to be highly
nonlocal in space, and often it is not clear, where the physical
processes responsible for the outcome of an experiment took place, in
the `device', the `contacts', or all over the place.

\begin{figure}
        \centerline{\epsfxsize=3in\epsfbox{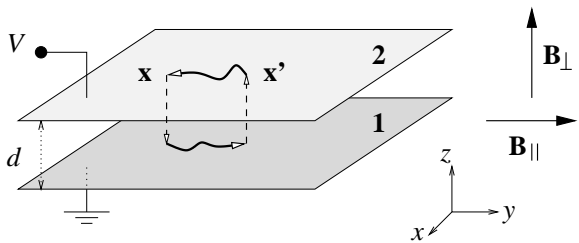}}

\vspace*{0.3cm}

        \caption{\label{setup}Schematic setup of the approach: two
          parallel two-dimensional electron systems are subject to a
          bias voltage and external magnetic fields. A tunneling current
          flowing between the layers is recorded as a function of the
          external control parameters.}
\end{figure}

These problems with contacted systems and local tunneling devices led
to the idea to use a second electron system with essentially known
properties as an {\em extended} tunneling spectroscope~\cite{S89}.
Setups of such type can conveniently be realized in double-quantum-well structures embedded in semiconductor
heterostructures~\cite{Iav-exp}, as shown schematically in
Fig.~\ref{setup}: Two electron systems confined in parallel wells
separated by an isolating barrier of uniform thickness $d\sim 10-20$ nm
form a double-layer system of two two-dimensional electron systems. A
tunneling current $I$ from one layer to the other is driven by
applying a voltage difference $V$ between them.  The tunneling region
is typically a few $100 \times 100$ $\mu$m$^2$ in extent.  In the
absence of any disorder scattering and/or tunneling amplitude
inhomogeneities, the tunneling from one layer to the other can only
occur if energy and momentum are conserved. This leads to the resonant
behavior of the tunneling current that is characteristic of
two-dimensional systems.  More generally, for constrained geometries
(e.g., quantum wire/two-dimensional electron system) the field
dependence of these resonances can be analyzed to obtain information
on the dispersion of the fundamental excitations in the two
systems~\cite{Luttinger}.

However, in `real' systems, inhomogeneities in the tunneling
barrier thickness, static disorder, and other non-momentum conserving
imperfections will lead to modifications of the idealized resonant
current profile. Of these intruding mechanisms, the first appears to
be the most serious: the current will respond with exponential
sensitivity to any fluctuations of the layer separation; for strong
enough spatial variations one may run into a scenario, where tunneling
occurs only at a sparse set of `hot spots', with no traces of a
resonant profile left\cite{OrHa}. (Some characteristics of this type of current
flow will be discussed below.)  However, recent technological advances
have made it possible to manufacture double-well systems with near-monolayer
precision. In such devices, fluctuations in the tunneling matrix
elements are reduced down to values of ${\cal O}(10 \%)$~\cite{cwhbarnes}
and can be absorbed into a renormalization of the effective in-plane
disorder. In the present paper, the focus will be on transport in these
near-planar devices.

Even if the tunneling is homogeneous, static disorder will
broaden the resonant behavior and introduce fluctuations.  The
broadening of the average current is related to the dynamics on short
time scales~\cite{Iav-theo}. In contrast, the fluctuations contain
information about physical processes on much larger time
scales~\cite{MAJ-annalen} of the order of, e.g., the diffusion time
through the system. It is the purpose of this paper to investigate the
nature of these fluctuations and their relation to the aforementioned
electronic correlation functions.

In fact, we will see below that detailed information on the
correlation functions $F^{[d;D;C]}$ can be extracted from the
tunneling current fluctuations without disturbing the system.
Moreover, (i) the tunneling takes place uniformly at all points of the
layers which means that an averaging over spatially fluctuating
structures (e.g.,  details of the microscopic wave function amplitudes)
is intrinsic to the data contained in the current. (ii) Several
parameters can be tuned to gain information: The bias voltage resolves
energetic correlations, a parallel magnetic field resolves spatial
correlations, and a perpendicular magnetic field may serve as a
control parameter for parametric correlations (i.e., correlations
between Green functions evaluated at different values of external
control parameters).  (iii) The geometry of
the layers can be designed freely, so that it is possible to study
different regimes of particle dynamics (e.g., ergodic, ballistic,
diffusive, etc.).

In this work, after the introduction of the general theoretical
background of the tunneling current statistics, we will consider two
prototypical system classes. First, we study extended systems for
which the phase coherence length $L_\phi$ is much larger than the
microscopic length $l_{\rm min}$ and a crossover or a transition from
diffusive motion to Anderson localization may take place. For such
systems, a parallel magnetic field $\B_\parallel$ can be employed as
an instrument for resolving the long-range behavior of the
correlation functions $F^{[D;C]}$.  Importantly, the field alignment
parallel to the two-dimensional planes implies that the charge carrier
dynamics is not affected by $\B_\parallel$.  While in the diffusive
regime explicit expressions for the correlation functions are known,
no quantitative expressions for regimes with strong (nonperturbative)
localization and interaction effects are available. However, in a
regime of localization-delocalization crossover or transition, scaling
behavior is expected to restrict the functional form of $F^{[D;C]}$.
This opens the possibility to extract the relevant critical indices
from tunneling conductance measurements.

Second, we study a geometry, where one of the layers forms a
ballistic quantum dot ($L<l_{\rm min}$) in the ergodic regime. The
other, extended, layer serves as the spectrometer. For the quantum dot,
the parametric correlations with respect to a perpendicular magnetic
field, present in $F^{[d;D;C]}$, can be obtained from the current
fluctuations.  A similar setup has already been realized
experimentally by Sivan et al.~\cite{Sivan94}. In that work, a single
level (in contrast to our extended system) was used as a spectrometer
to study a quantum dot device.  This experiment led to results for the
functional form of the correlator $F^{[d]}$, compatible with theoretical
predictions from random matrix theory. However, one would expect that
the data obtained from single-level spectroscopy is still weighted
with nonuniversal wave-function amplitudes specific to the isolated
`monitor level'. In contrast, for the two-dimensional layer/quantum
dot setup considered here, the current flow is extended and spatially
uniform. As a consequence, the tunneling current fluctuations are
microscopically related to the purely spectral content of parametric
correlations. Below we will establish the quantitative connection
between the field and voltage dependence of the tunneling current
fluctuations and a number of correlation functions that have been
analyzed in the recent theoretical literature~\cite{paracorr}. Moreover,
we will try to assess to what extent these connections, obtained for
the chaotic noninteracting electron gas, may be susceptible to
interaction mechanisms such as Coulomb drag\cite{ZMcD93b,Rojo} or Coulomb blockade effects~\cite{CouBloc}.

 The outline of the paper is as follows: In Sec.~\ref{sec-theo}
we specify our model and review the general formula for the tunneling
current.  Turning to new results, we show that the Fourier transform
of the tunneling conductance correlator with respect to the magnetic
field directly yields the spatially resolved correlation functions
$F^{[D;C]}$.  In Sec.~\ref{sec-crit} the theory will be
applied to diffusive and anomalous diffusive systems, and in
Sec.~\ref{sec-spectral} to an analysis of spectral and parametric
correlations in finite quantum dots.  The impact of Coulomb charging
effects on the tunneling current fluctuations will be discussed in
Sec.~\ref{sec-spectral}. We conclude in Sec.~\ref{sec-conc}.

\section{Theory of tunneling currents}\label{sec-theo}

\subsection{The current formula}\label{sub-currformula}

Consider a double-layer system consisting of two parallel
two-dimensional electron gases (2DEGs), labelled 1 and 2, respectively (see
Fig.~\ref{setup}).  The two layers are separated by a tunneling
barrier that we assume to be uniform.  We aim to analyze the
tunneling current $I$ under conditions, where the tunneling is weak (in
a sense to be specified momentarily).  After matching the electron densities
in both layers by adjusting their chemical potential $\mu$, the
current becomes a
function of bias voltage $V$, temperature $T$, and, optionally, a
magnetic field $\B$, $I=I(V,\B;T,\mu)$.

Quantum mechanically, the system can be represented in terms of a tunneling
Hamiltonian, $H = H_1 + H_2 + H_{\cal T}$, where $H_1$ and $H_2$ describe layer 1 and 2,
respectively, while $H_{\cal T}$ describes the transfer of
electrons between the layers~\cite{H_tunnel}. Choosing a gauge where the bias voltage
has been transferred to the tunneling matrix elements, $H_{\cal T}$
can be written as
\begin{eqnarray}
H_{\cal T} = \int\!
d^2x\, d^2x' \LK{\cal T}_{\x\x'}
e^{-iVt}{\psi^\dagger}_1(\x)\psi_2(\x')+{\rm h.c.} \RK \,, \label{E2}
\end{eqnarray}
where ${\cal T}_{\x\x'}$ is the tunneling amplitude from $\x$ in layer
1 to $\x'$ in layer 2, and $\psi^\dagger_j$, $\psi_j$ are electron
creation and annihilation fields for layer $j=1,2$. For convenience, we
use units where $\hbar\!=\!c\!=\!k_{\rm B}\!=\!e\!=\!1$.  
Since the tunneling matrix elements ${\cal T}_{\x\x'}$ decrease
exponentially (on atomic scales) as a function of $|\x -\x'|$ we model
${\cal T}_{\x\x'}$ as a spatially local object, ${\cal T}_{\x
  \x'}={\cal T}_{\x }\delta(\x-\x')$.
In the {\em weak tunneling
  regime}, i.e., when the typical time after which an
electron tunnels is larger than the characteristic time scale that is
to be resolved in the experiment, single-tunneling events dominate~\cite{fn_tunneling}. Then, the tunneling current
reads~\cite{Mahan,Iav-theo}
\begin{eqnarray}
I(V,\B) &=& 2 \int\! d^2x \, d^2x'
\int\!(d\epsilon^{[V]}) \, {\cal T}_{{\x}}{\cal T}^*_{{\x'}}\,
e^{{i}{\q}({\x}-{\x'})}\times \nonumber\\
&& \times\, A_1({\x},{\x'};\epsilon,{B_\perp})
A_2({\x'},{\x};\epsilon-V,{B_\perp}) \,, \label{current}
\end{eqnarray}
with the abbreviation $\int(d\epsilon^{[V]})=\int_{-\infty}^{\infty}
\!  \frac{d\epsilon}{2\pi} [ n_{\rm F}(\epsilon-V)-n_{\rm
  F}(\epsilon)]$.  Here $n_{\rm
  F}(\epsilon)=(1+e^{(\epsilon-\epsilon_{\rm F})/T})^{-1}$ is the
Fermi distribution function at temperature $T$ and Fermi energy
$\epsilon_{\rm F}$. The characteristic momentum scale set by the parallel field is $\q= d \B_\parallel \times {\bf e}_z$,
where ${\bf e}_z$ is the unit vector perpendicular to the plane.

In Eq.~(\ref{current}) the quantities of main interest are the spectral functions of layer $j$, $A_j= i
(G_j^+-G_j^-)$, where $G_j^\pm$
is the retarded/advanced one-particle Green function. 
These spectral functions depend on the bias voltage $V$ applied to layer
2, and on the perpendicular magnetic field ${B_\perp}$.  In fact, it
will be our main objective to obtain information on these objects {\em
  through} their parameter dependences.  In this context, it is
crucial to note that $\B_\parallel$ does not change the dynamics
within the individual layers, but merely weighs the tunneling current
with an Aharonov-Bohm-type phase. The sensitivity of the current to
this flux will help to gain information about the long-range
propagation within the layers.  A caricature of the basic idea is
depicted in Fig.~\ref{setup}. This figure illustrates the basic
physical processes underlying the current flow as described by
Eq.~(\ref{current}): An electron tunnels at point ${\x}$ from layer 2
to 1, leaving a hole behind.  It propagates within that layer to point
${\x'}$, where it tunnels back to layer 2 and recombines with the
hole.  The in-plane magnetic field enters the formula via the flux
through this electron-hole loop. Therefore, the in-plane magnetic
field dependence of the tunneling current contains information about
the typical area enclosed in the loop that in turn is determined by
the typical range of propagation within the layers.  

To further simplify the analysis, we note that in the absence of significant
interaction corrections the spectral functions themselves do not exhibit
temperature dependence. 
Under these conditions, a simple integral
relation between currents at zero and finite temperatures holds:
\begin{eqnarray}
I(T,\epsilon_{\rm F}) = -\int\!d\epsilon\, \LK
\frac{\partial n_{\rm F}}{\partial\epsilon}\RK I(T=0,\epsilon_{\rm
  F}=\epsilon) \,. \label{E6}
\end{eqnarray}
In the following, unless stated otherwise, all results will be given
for $T=0$ only. The generalization to finite temperature --
essentially a smearing of the $T=0$ results -- obtains from Eq.~(\ref{E6}).

In this paper, we are primarily interested in the tunneling current
flowing between disordered systems. The microscopic properties of
these systems will be described by some disorder distribution function
about which we make three idealizing assumptions:

 (1) The disorder
potentials of the two layers are essentially
uncorrelated~\cite{fn-dis}.  

(2) The e-e interaction and higher order
tunneling processes are not able to introduce significant interlayer
correlations in the motion of the charge carriers. Practically, this
means that impurity averages can be taken for each layer
independently. Roughly speaking, e-e interaction effects can be
divided into three groups: momentum transfer between the layers
(`Coulomb drag'), charging effects associated with the tunneling
process, and self-energy corrections (due to inelastic scattering and
dephasing). Sizable interlayer Coulomb correlations may arise in very
clean systems subject to strong perpendicular magnetic fields.  In
such systems, the e-e interaction can stabilize a fractional quantum
Hall phase~\cite{PrangeGirvin} and interlayer e-e interactions lead to
additional correlation phenomena (spontaneous coherence and quantum
Hall ferromagnets~\cite{Girvin-LesHouches}). In contrast, for strong
enough disorder, the e-e interaction is less
significant~\cite{fn-drag} and, thus, the effective random potentials
in each layer can be treated as statistically independent of each
other. This assumption can be tested experimentally, as will be
discussed below.  Charging effects will inevitably influence the
tunneling current at low bias; we postpone a discussion of these
corrections until Sec.~\ref{sub-avcur}. 

(3) Disorder does not
significantly affect the spatial homogeneity of the tunneling, i.e.
the tunneling matrix elements do not depend on position, ${\cal
  T}_\x\equiv{\cal T}$. The spatial homogeneity of the tunneling is
very sensitive to the thickness of the barrier.  However, it is now
possible to grow heterostructures with near-monolayer precision and,
thus, achieve a tunneling probability that is almost spatially
constant~\cite{Eis95,Tur96}. (In high precision devices, the space
dependent relative fluctuations in the tunneling probability can be
reduced to about 10\% and lower~\cite{cwhbarnes}.) The
validity of this assumption can be tested experimentally.  In the
extreme case, where tunneling occurs only through tunneling centers or
`pinholes'~\cite{OrHa}, the resonant behavior of the tunneling
current disappears. If the spacing between pinholes exceeds the mean
free path, the average current is just proportional to the product of
the local densities of states, $\langle I\rangle\sim\nu_1\nu_2$, and,
furthermore, becomes independent of magnetic field. 

In principle, a
full (angle-resolved) analysis of the field dependence of the current
would allow one to extract information about the distribution of
tunneling centers~\cite{levitov}.  However, further discussion of this
type of `tunneling center spectroscopy' is beyond the scope of this
paper. Keeping in mind that the working assumption of near-homogeneous
tunneling can be put to experimental test, we hereafter concentrate on
the case ${\cal
  T}_\x\equiv{\cal T}$. (Note that, although significant inhomogeneities
would largely
obstruct the detection of transport correlations, they only have a
minor effect on the analysis of {\em spectral} correlations.)

\subsection{Average current}\label{sub-avcur}

To foster the discussion of fluctuations below, this section briefly
recapitulates the main characteristics of the average tunnel current~\cite{fn-av}.
Under the assumptions formulated above it is
given by
\begin{eqnarray}
\langle I(V,{\B})\rangle&=&\frac{L^2}{\pi}\left|{\cal T}\right|^2 \int\limits_0^{V} \!
         d\epsilon \int\! d^2x
        \, e^{i\q\x}\times\nonumber\\
&& \times\langle A_1(\x;\epsilon,B_\perp)\rangle
         \langle A_2(-\x;\epsilon-V,B_\perp)\rangle\,.\label{av}
\end{eqnarray}
The current, Eq.~(\ref{av}), is characterized by the averaged one-particle
Green function $\langle {G^\pm}_j(\x,\x';\epsilon) \rangle$. This
quantity is short-ranged on a scale $l_{\rm min}$. For small
perpendicular magnetic fields $B_\perp$, $l_{\rm min}$ is set by the
mean free path $l=v_{\rm F} \tau$, where $v_{\rm F}$ is the Fermi
velocity and $\tau$ the elastic scattering time. (In cases, where the
scattering times in the layers are different, we need to generalize to
$\tau_j$, $j=1,2$.)  We here focus on
systems, where the mean free path $l$ is much shorter than the linear system size $L$~\cite{fn-ltr}.
For stronger perpendicular magnetic fields, with cyclotron frequency
$\omega_c=B_\perp/m$ exceeding the inverse elastic scattering rate $\tau^{-1}$, the classical cyclotron radius $R_c=v_{\rm F}/\omega_c$
is smaller than the mean free path $l$ implying that $l_{\rm
  min}=R_c$ (see, e.g., Ref.~\cite{Stone}).

The average current has been studied theoretically~\cite{Iav-theo}
and experimentally~\cite{Eis95,Tur96}. For matched Fermi energies and vanishing magnetic
fields, the theoretical result reads
\begin{eqnarray}
        \frac{\langle I (V,\B=0)\rangle}{V} =
        G_0 \frac{\Gamma^2}{V^2+\Gamma^2} \,,\qquad G_0 = \frac{2\nu|{\cal T}|^2}{\Gamma}L^2\,,
        \label{E7}
\end{eqnarray}
where $\Gamma=\Gamma_1+\Gamma_2$, $\Gamma_j=1/(2\tau_j)$ is the
line-width of the Lorentzian-shaped average spectral function in layer
$j$, $\nu$ is the single-particle-level density of states per unit
area, and $G_0$ is
the characteristic low-bias average conductance of the system.
For the average current, the
condition of weak tunneling is
\begin{eqnarray}
\Gamma_{\cal T} < \Gamma \,, \label{E9}
\end{eqnarray}
where $\Gamma_{\cal T}\equiv \left| {\cal T}\right|^2 \Gamma^{-1}$
is the inverse of the (golden rule) rate at which a particle
propagating in one layer tunnels to the other. Henceforth we will focus on the regime of small bias
voltage $V \ll \Gamma$. Under these conditions,
$\langle I (V,\B=0)\rangle/V\approx G_0$.
In the presence of a moderately weak in-plane magnetic field ($|\q|^{-1}\gg$ Fermi
wavelength), this generalizes to
\begin{eqnarray}
 \frac{\BRA
  I(V,\B)\KET}{V} = G_0 f(|\q|l) \,, \label{E10}
\end{eqnarray}
where the scaling function $f$ exhibits the asymptotic behavior
$f(x\ll 1) =
1+ {\cal{O}}(x^2)$ and $f(x\gg 1)\sim x^{-1}$. In the following, we
will concentrate on the weak field regime,
$|\q|l\ll 1\Rightarrow f\approx 1$.

Before moving on to our main issue, mesoscopic fluctuations of
$I(V,\B)$, let us make a few more remarks on the average current.
Specifically, we wish to argue that from the known behavior of
$\langle I \rangle$, conclusions on the validity of some of the
assumptions made above can be drawn.  In Refs.~\cite{Eis95} and \cite{Tur96} the
influence of a perpendicular magnetic field on the average current
between high mobility samples was investigated. It was found that
strong fields lead to a suppression of the differential tunneling
conductance, $G_{\cal T}\equiv \partial I/\partial V$, at zero bias.  This
phenomenon is called the `tunneling-gap'. The splitting of the
conductance peak at $V=0$, characterized by some field-dependent
gap energy $\Delta(B)$, is due to the Coulomb interaction within and
between the layers.  As shown theoretically in~\cite{LevShy95-b} and
confirmed experimentally in Refs.~\cite{Eis95} and \cite{Tur96}, the total current
flowing at the split peaks (positioned at $V=\pm \Delta$) equals the
peak current at $V=0$ in the absence of interactions. This observation
implies that interactions in these experiments largely manifest
themselves in the form of self-energy corrections to the single-particle poles. In contrast, if strong interlayer correlation effects
were present, the peak current would increase.  Indeed, for strong
interlayer correlations, the momenta of the particle and the hole
constituting the `current-loop' would be partially correlated. This
should lead to gradual resurrection of the resonant behavior
characteristic for the clean case and, therefore, to an unsplit zero-bias conductance peak. 

Even at $B\!=\!0$ the diffusive zero-bias anomaly leads
to some splitting of the peak. However, the separation of the maxima
of the subpeaks is small in $1/\sqrt g$~\cite{zero-bias} and will be
neglected here. Note that, when lowering $g$, nonperturbative (in
$1/g$) zero-bias effects can occur~\cite{grabert}. Finally, in
Sec.~\ref{QD-C} the analog of the zero-bias anomaly in finite quantum
dots (Coulomb blockade) will be considered.

\subsection{Fluctuations}\label{sec-fluct}

We next turn our attention to the fluctuations of the tunneling current.
As a starting point we will use the formula
\begin{eqnarray}
I(V,\B) &=& \frac1\pi |{\cal T}|^2 \int\! d^2x \, d^2x'
\int \limits_{\epsilon_{\rm F}}^{\epsilon_{\rm F}+V}\!d\epsilon \,
e^{{i}{\q}({\x}-{\x'})}\times \nonumber\\
&& \times\, A_1({\x},{\x'};\epsilon,B_\perp)
A_2({\x'},{\x};\epsilon-V,B_\perp)\label{current_uni}
\end{eqnarray}
for the zero-temperature current at uniform tunneling probability.
As we are interested in correlations on large time scales,
Eq.~(\ref{E9}) for the range of applicability of the weak tunneling
approximation has to be replaced by the more restrictive condition
\begin{eqnarray}
\Gamma_{\cal T} < {\rm max\,}(\tau_\phi^{-1},V,v_{\rm F} |\q|) \,,\label{E9b}
\end{eqnarray}
where $\tau_\phi$ is the phase coherence time and $v_{\rm F}|\q|$ the
characteristic energy scale set by the parallel magnetic field. This
inequality states that the probability for an electron to tunnel,
while moving coherently within one layer, is low.

To describe fluctuations of the current and related quantities, we
will consider correlation functions  of the type
$$
C_X(z,z') \equiv {\langle X(z) X(z')\rangle_c\over \langle X
  \rangle^2},
$$
where $X$ is an observable, $\langle AB \rangle_c \equiv \langle AB
\rangle - \langle A \rangle \langle B \rangle$, and $z^{(\prime)}$
represents the set of parameters $\{V^{(\prime)}, \B^{(\prime)},
\epsilon_{\rm F}^{(\prime)}\}$. [To avoid confusion, let us
reiterate that in this paper angular brackets stand for averaging
over an external set of parameters, {\em not} for a quantum mechanical
average. For example, in our discussion below, $X(z) \equiv G_{\cal T}(V,{\bf
  B})$ may stand for the conductance measured at a certain
field/voltage configuration. The subsequent $\langle
\dots\rangle$-average will then be over configurational fluctuations.]
The suppression of the parameter dependence in the normalization
denominator indicates that, on the $z$-scales relevant for the
structure of fluctuations, the parameter dependence of the averaged
observables is negligibly weak.

In most of the following, we will concentrate on the correlation
function $C_G$ of the differential conductance $G_{\cal T}$. This
quantity is (a) experimentally more relevant than the current
correlation function $C_I$ and (b) tends to exhibit more pronounced
structure. Indeed, it is straightforward to show that the current and
the conductance correlation function, respectively, are related
through $C_I(V) = 2V^{-2}\int_0^{V} d\omega \LK V- \omega \RK
C_G(\omega)$, i.e. $C_I$ is obtained from $C_G$ through an integral
average.

Averaging (the square of) Eq.~(\ref{current_uni}) over disorder, one
verifies that
\begin{eqnarray}
&&         C_G(\omega,\B,\B') =\frac1{G_0^2}\langle G_{\cal T}(V;\B)G_{\cal T}(V';\B')\rangle=\nonumber\\
&&\quad= \frac{4|{\cal T}|^4}{\pi^2G_0^2} \int \prod_{n=1}^4 d^2x_n \,
    e^{i\q (\x_1-\x_2)+ i\q'(\x_3-\x_4)}\times \nonumber\\
&&\qquad\times \Re F_1(\x_1,\x_2,\x_3,\x_4;z)
    \, \Re F_2(\x_2,\x_1,\x_4,\x_3;z) \,, \label{E24}
\end{eqnarray}
where $G_0 = \langle G_{\cal T} \rangle$ as discussed in the
previous section.
The objects
\begin{eqnarray*}
&&F_{1/2}(\x_1,\x_2,\x_3,\x_4;z) =\\
&&\qquad =\langle G_{1/2}^-(\x_1,\x_2;\epsilon,B_\perp)
G_{1/2}^+(\x_3,\x_4;\epsilon+\omega,B_\perp^\prime)\rangle\,,
\end{eqnarray*}
where $z=\{\omega,B_\perp,B_\perp'\}$, and $\omega=V-V'$, are the
basic two-particle correlation functions discussed in the
Introduction.  That Eq.~(\ref{E24}) contains the product of two of
these correlators is a direct consequence of our assumption of
negligible interlayer disorder correlation. Notice that while the
correlation functions $F$ of noninteracting systems categorically
depend only on the energy difference between the two Green functions,
the dependence on the perpendicular fields can be more complicated.

\begin{figure}
\centerline{\epsfxsize=3in\epsfbox{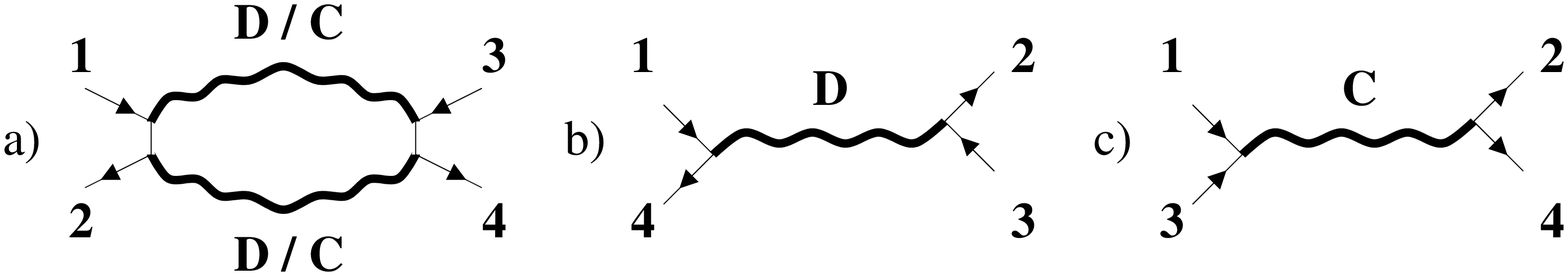}}

\vspace*{0.3cm}

\caption{\label{diag}Diagrams contributing to $C_{G/I}$: a) density-density, b) diffuson,
and c) cooperon. The thick wavy lines denote the diffuson and cooperon propagators.}
\end{figure}

The fourfold integration over the coordinates $\x_i$
implies that all three contributions discussed above, density-density
$F^{[d]}$, diffuson $F^{[D]}$, and cooperon $F^{[C]}$, contribute to
Eq.~(\ref{E24}) (see Fig.~\ref{diag}).
At this stage, the role of the weak in-plane magnetic field becomes
clear. As discussed above, the correlators $F^{[d;D;C]}$ are long ranged (as compared to the microscopic spatial extent of the average
Green functions contributing to $\langle I \rangle$). This means that
Eq.~(\ref{E24}) is field sensitive -- through the magnetic wave vector
-- on {\em small} magnetic field scales.  The characteristic field
strength is determined through $|\q| = d B_\parallel \sim
L_\omega^{-1}$, where $L_\omega$ is the typical distance a particle
propagates during time $\omega^{-1}$. For example, for a medium characterized
by diffusive motion with diffusion constant $D$, $L_\omega \sim
(D/\omega)^{1/2}$. Using that for the three fundamental correlators
the coordinates are pairwise equal [with an accuracy of ${\cal
  O}(l_{\rm min})$]
and neglecting factors $\sim |\q| l_{\rm min} \ll 1$, Eq.~(\ref{E24}) assumes the
form
\begin{eqnarray}
&&C_G(\omega,\B,\B') =\left(\frac{8\pi l^2 |{\cal T}|^2}{g G_0}\right)^2
\int d^2(x,x') \times \nonumber\\
&&\qquad\hspace{-0.3cm}\times \Big(  e^{i(\q-\q')(\x-\x')}
\;\Re F^{[D]}_1(\x,\x';z)
    \, \Re F^{[D]}_2(\x',\x;z) +\nonumber\\
&&\qquad  +e^{i(\q+\q')(\x-\x')}
\;\Re F^{[C]}_1(\x,\x';z)
    \, \Re F^{[C]}_2(\x',\x;z) +\nonumber\\
&&\qquad
+\Re F^{[d]}_1(\x,\x';z)
    \, \Re F^{[d]}_2(\x',\x;z)\Big)\,,
\label{E24a}
\end{eqnarray}
where $g=2\pi D\nu$ is the dimensionless conductance.

Equation (\ref{E24a}) states that the diffuson contribution
$F^{[D]}$ couples to the difference, $\B^-_\parallel \equiv
\B_\parallel - \B_\parallel'$, of the two in-plane field vectors, the
cooperon contribution $F^{[C]}$ to the sum, $\B^+_\parallel \equiv
\B_\parallel + \B_\parallel'$, whereas the density-density
contribution $F^{[d]}$ is $\B_\parallel$-insensitive. (Later on we will see how
information on $F^{[d]}$ can be extracted from the dependence on the
perpendicular field $B_\perp$.) Equation (\ref{E24a}) holds true for extended systems, where the unconstrained integration over $x,x'$ implies momentum conservation, as well as for restricted systems, where the in-plane momentum is not conserved in tunneling.

Finally, if both systems are extended, Fourier transforming Eq.~(\ref{E24a}) in the magnetic field we obtain
\begin{eqnarray}
&&\Re F^{[D]}_1(\Delta \x;z)
    \, \Re F^{[D]}_2(-\Delta \x;z) = 
\left(\frac{\nu^2dL}{8\pi}\right)^2\times\nonumber\\
&&\qquad\int d^2 B_\parallel^+ e^{-i d
      (\Delta \x \times \B_\parallel^+)_z}
    C_G(\omega,\B^+,\B^-),\nonumber\\
&&\Re F^{[C]}_1(\Delta \x;z)
    \, \Re F^{[C]}_2(-\Delta \x;z) = 
\left(\frac{\nu^2dL}{8\pi}\right)^2\times\nonumber\\
&&\qquad\int d^2 B_\parallel^- e^{-i d
      (\Delta \x \times \B_\parallel^-)_z}
    C_G(\omega,\B^+,\B^-),
\label{central}
\end{eqnarray}
where $\Delta \x$ is the difference between the two spatial arguments
of the correlation functions $F$, 
and we have used that $\q \x = d (\x \times \B_\parallel)_z$ as well
as the result~(\ref{E7}) for $G_0$.

Equation (\ref{central}) contains a central message of our paper: Detailed
spectral and spatial information on the correlation functions $F$ can
be obtained from the dependence of the tunneling current on a parallel
magnetic field.  (In contrast to contact measurements,) the current
approach to exploring correlation functions enables one to
continuously measure spatial scale dependences, and does not
incorporate strong local perturbations.  If one of the layers is a finite quantum dot, Eq.~(\ref{E24a}) gives the general relation between the current fluctuations and the spectral correlation functions. In the next two sections, we
will discuss applications of this general concept to some concrete problems.

\section{Anomalous diffusion}\label{sec-crit}

In this section, we are going to apply Eq.~(\ref{central}) to the
problem of (anomalous) diffusion in spatially extended structures. We
first note that for the limiting cases of purely ballistic and
diffusive dynamics, respectively, the correlation functions $F^{[D;C]}$ can
be calculated explicitly. For ballistic systems, a straightforward
integration over the momenta of the single particle Green functions
obtains
\begin{eqnarray}
 \Re
F^{[D;C]} (r;\omega) &\sim& \frac{\nu}{v_{\rm F}r} \cos\frac{\omega
  r}{v_{\rm F}}\,. \label{E30}
\end{eqnarray}
For diffusive systems, leading-order diagrammatic perturbation theory
(one diffuson/cooperon approximation) leads to
\begin{eqnarray}
\Re F^{[D;C]}
(r;\omega) &\sim& \frac{\nu^2}g {\rm
  ker}\left(\frac{r}{L_\omega}\right) + \dots\,, \label{E31}
\end{eqnarray}
where ${\rm ker}(x)$ is the Thomson function~\cite{grad}. For
small $x$, this function can be  approximated as ${\rm ker}(x)\approx - C -
\ln(x/2)$ ($C\approx 0.58$ Euler's constant). The ellipses stand for weak-localization-type contributions of
higher order in the number of diffusons and cooperons. These
corrections scale with negative powers of the dimensionless
conductance $g$. By definition, we will denote a system as
`diffusive' if $g\gg 1$ and weak localization does not play a role.

To get some idea about the strength  of the tunneling current
fluctuations let us briefly discuss the $\B=0$ current correlation
function $C_I(V)$  for two different setups: (a) two disordered
layers and (b) only one disordered layer and one `clean' layer.
Here `clean' means that $l$ is much larger than system size $L$.
Substituting Eqs.~(\ref{E30}) and (\ref{E31}) into Eq.~(\ref{E24})
and integrating over frequencies one finds in case (a)
\begin{eqnarray}
C_I(V) = g^{-2}
\frac{E_{\rm Th}}{V}[ c^{[d]} + 2c^{[D]} \ln(V/E_{\rm Th})]
\,,\label{31}
\end{eqnarray}
where $E_{\rm Th}=g\delta$ is the Thouless energy. Eq.~(\ref{31})
has been derived under the assumption $V> E_{\rm Th}$. Physically,
this means that on time scales $t\sim V^{-1}$, the charge carriers
do not have enough time to diffusively explore the entire system
area. For smaller voltages, a crossover to an ergodic regime,
discussed in the next section, takes place.  The two numerical
coefficients $c^{[d]}=9/(8\pi)$ and $c^{[D]}=1/(4\pi)$ determine
the strength of the density-density and diffuson contribution,
respectively. The factor of 2 multiplying $c^{[D]}$ expresses
the fact that in the field-free case, the diffuson and cooperon
contribution, respectively, are equal and add. For case (b), the
expression looks similar, however, instead of the logarithm a
factor $\sqrt{V/\Gamma}$ appears. This means that in the regime of
interest, $V\ll \Gamma$, the current fluctuations between a clean
and a disordered layer are largely due to fluctuations in the
density of states. This result can easily be understood
qualitatively: In a clean system, the charge carriers move much
faster than in a disordered system. As a result, the particles
propagating in the disordered system do not have enough
time to diffusively travel over large distances. This in turn
implies that the diffuson and cooperon contribution to the
correlation are reduced by a phase space reduction factor, whereas
the density-density contribution, involving only Green functions taken at coinciding points (within the clean system), remains
unaffected.  Actually, the density-density contribution to the current correlation is proportional to the variance of the number of levels in an energy
window of $V$. This is very similar to the conductance fluctuations in
conventional transport,
which are related to the level number variance in an energy window of the
size of $E_{\rm Th}$. For completeness, we mention that for case (b),
$c^{[d]}=4/\pi$ and $c^{[D]}=16/(3\pi)$.  Note that the
conductance is self-averaging ($\sim L^{-2})$ in the thermodynamic
limit. Furthermore, the fluctuations are suppressed by a factor
$g^{-2}$. This is a phase-space reduction factor expressing the
fact that to obtain averaging-insensitive contributions, two of
the four spatial arguments of the correlation function $F$ must be
close to each other on scales $l$, cf.  Eq.~(\ref{E24a}). Finally,
we notice that already weak {\em perpendicular} magnetic fields of
${\cal O}(1/\Delta x^2)$, where $\Delta x^2$ is the characteristic
area of extent of $F^{[C]}$, suffice to suppress the cooperon
contribution.  This means that the dependence of the current
fluctuations on a perpendicular field can be used to determine the
maximum range of the correlation functions at frequencies $\omega
\sim V$. For $V < \tau_\phi^{-1}$, this scale is set by the
dephasing length, $L_\phi$.  In analogy to the classical
experiments by Bergmann~\cite{Bergmann}, the field dependence of
the current for these low voltages can be used to estimate
$L_\phi$.

What can be said about systems with more complex types of dynamics,
i.e., systems where localization and/or interaction corrections play
some role.  In principle, both weak localization and interaction
corrections can be taken into account perturbatively, where the
inverse of the dimensionless conductance, $g^{-1}$, represents the
expansion parameter~\cite{zero-bias}. As $g$ is lowered, these
nondiffusive corrections become stronger and eventually, for $g=
{\cal O}(1)$, the perturbative description breaks down. However,
relying on concepts of scaling theory, it is still possible to make
some general statements about the behavior of the strongly disordered
electron gas: For $g\sim1$, localization phenomena begin to
qualitatively affect the dynamics.  According to
the one-parameter scaling theory~\cite{Abrahams79}, the weakly
interacting electron gas eventually flows into a localized regime
provided that (a) spin-orbit scattering is negligible and (b) no
strong perpendicular magnetic fields are present. In contrast, systems
with significant spin-orbit scattering  are expected to exhibit a true
metal-insulator transition at some critical value $g_c\sim {\cal
  O}(1)$~\cite{Mer98}; 2D electrons (interacting as well as
noninteracting) subject to strong magnetic fields undergo a
metal-insulator transition responsible for the quantum Hall
effect~\cite{IQHEBook94}. Finally, in a number of experiments on 2D
electrons with strong interaction parameter $r_s>1$ transport
behavior has been observed that resembles a metal-insulator
transition~\cite{2dmetalinsulator}, too.

In all these phenomena (except, perhaps, the not sufficiently well-understood transport phenomena discussed in Ref.~\cite{2dmetalinsulator}) the concept of `anomalous diffusion' plays a
key role~\cite{Chalker90}. Prior to the onset of strong localization,
the electron dynamics undergoes a crossover from ordinary diffusive
($g \gg 1$) to anomalously diffusive ($g\sim 1$).
Quite generally, the correlation
function of anomalously diffusive electrons has the scaling form
\begin{eqnarray}
F^{[D]}(r,\omega) \sim \LK \frac{r}{L_\omega}\RK^{-\eta} e^{-2r/\xi}\,,
\label{scaling}
\end{eqnarray}
where $\xi$ is the localization length and $\eta$ a characteristic
exponent related to the multifractal nature of states that are
neither regularly extended nor fully localized~\cite{review}. The length $L_\omega$ is related to the energy $\omega$ by the
so-called {\em dynamical exponent} $z$,
\begin{eqnarray}
L_\omega \sim \omega^{-1/z}.
\end{eqnarray}
Whereas for noninteracting systems $z=2$ as in ordinary diffusive
systems, the value of $z$ for interacting systems is
controversial~\cite{HucBac99}.  In systems with a true
localization-delocalization transition, the localization length
diverges with a characteristic exponent $\nu$ upon approaching the
transition point.  In Eq.~(\ref{scaling}), we have assumed that $\xi$
is smaller than $L_\omega$. In the opposite case, $L_\omega$
would be the scale of exponential decay of the correlation function.

According to Eq.~(\ref{scaling}), the `non-diffusivity' of the
electron dynamics can be characterized in terms of the three exponents
$z,\nu$, and $\eta$. To obtain these quantities one needs to know both
the spatial and the energetic profile of the correlation function. In
fact, the aforementioned difficulty to continuously monitor the
spatial structure of electronic correlations has prevented previous
experiments from determining the exponent $\eta$. In contrast, the
basic relation (\ref{central}) does, at least in principle, contain
all the information needed to extract all exponents of anomalous
diffusion. In the following, we shall try to assess whether this
approach might work {\em in practice}.

One aspect counteracting the application of the current approach to
the analysis of anomalous diffusion is that to date semiconductor
devices tend to be `too clean': In state-of-the-art high-mobility
samples, the mean free path is of the same order as the low-temperature phase coherence length, roughly about 10 $\mu$m.  In such
devices, the phase coherent electron transport is ballistic and not
even conventionally diffusive. We thus need to consider low-mobility
devices, where the disorder concentration is increased either by
doping or by lowering the separation between the 2DEG and the donor
impurities. We expect that by artificially increasing the disorder, an
order of magnitude separation between $l$ and $L_\phi$ might be
attainable~\cite{wieck}.  Second, to observe significant deviations
from standard diffusion, we need to be in a regime of a low global
conductance $g$.  In low-mobility systems showing integer quantum Hall
transitions (when placed in strong perpendicular magnetic fields), the
typical Coulomb energy is low as compared to the disorder energy
scale.  In such systems, the conductance $g$ is of order unity in the
transition regimes, and anomalous diffusion might be observable by our
method. Notice, however, that for $g={\cal O}(1)$, the
tunneling current (like any other mesoscopic transport observable)
will be subject to significant renormalization by interaction
processes. More specifically, when lowering $g$, one will run into a regime,
where the zero-bias anomaly renormalization of the tunneling DoS
ceases to be small. However, a quantitative analysis of the interplay
of anomalous diffusion and interaction is beyond the scope of the
present paper.

\section{Spectral and parametric correlations}\label{sec-spectral}

\begin{figure}
\centerline{\epsfxsize=3in\epsfbox{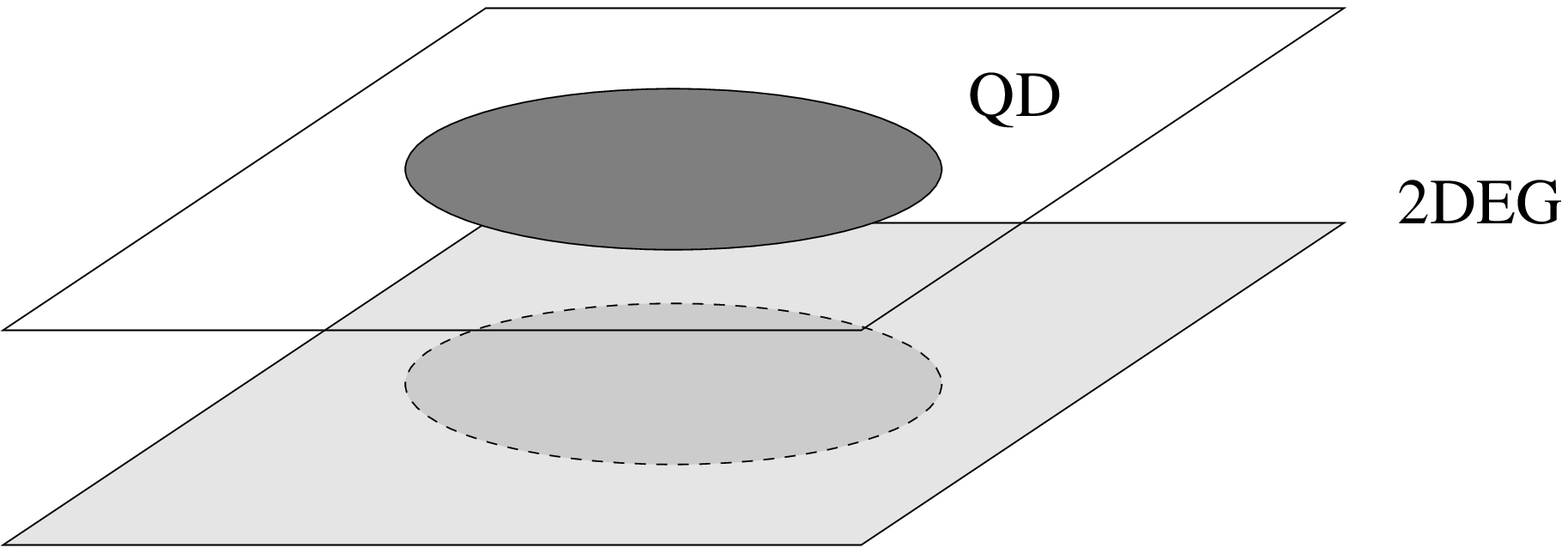}}

\vspace*{0.3cm}

\caption{\label{qd2deg}QD - 2DEG.}
\end{figure}

In the previous section, the focus was on exploring the spatial
structure of electronic correlation functions in extended systems.  We
now turn to a complementary application, viz., the analysis of {\em
  spectral} structures in finite chaotic environments. Specifically,
we shall consider a situation where one of the layers -- either by
top gates or through etching -- has been converted into a quantum
dot (QD) of characteristic size $L$, see Fig.~\ref{qd2deg}. The
2DEG underneath is extended as before.

Our objective is to analyze spectral and parametric correlations
through the tunneling current statistics. As before, the bias voltage
$V$ and the parallel magnetic field $\B_\parallel$ will be used to
identify different contributions to the current-current correlator,
and to detect their dependence. In addition to these control
parameters, we will employ a {\em perpendicular} magnetic field $B_\perp$ to
probe parametric correlations.

In the finite size setup under consideration, Coulomb charging effects
are likely to play some role. For didactical reasons, we will begin by
discussing the idealized noninteracting situation in
Sec.~\ref{QD-noC}. Coulomb corrections will then be considered
in Sec.~\ref{QD-C} where it will be argued that the impact of
interactions on the current fluctuations sensitively depends on the
parameter regime under consideration.

\subsection{Noninteracting case}\label{QD-noC}

We are interested in the behavior of a chaotic quantum dot on time
scales, where the electron dynamics is ergodic -- a `zero-dimensional'
system in the standard terminology of mesoscopic physics.  For a
diffusive system, the time it takes to establish ergodic dynamics is
set by $E_{\rm Th}^{-1}$. For nearly clean quantum dots, the
ergodicity time depends on the specifics of the boundary scattering
potential.  Ergodic mesoscopic systems are tailor made to modeling in
terms of random matrix theory (RMT). For simplicity, we shall assume
that the perpendicular magnetic field is strong enough to globally
break time reversal invariance. On the other hand, the field is
assumed to be too weak to significantly affect the ballistic dynamics
of the charge carriers: $R_c \gg {\rm \,min}(L,l)$. Under these
conditions, the quantum dot can be modeled in terms of random matrices
drawn from the Gaussian Unitary Ensemble (GUE).  Specifically, we will
describe the upper system in terms of some $N$-dimensional Hermitian
matrix Hamiltonian $(H+i\Phi)_{\mu\nu}$, where $H$ is taken from a GUE
and the antisymmetric matrix $\Phi$ represents the magnetic field in a
way to be discussed momentarily.  The indices $\mu,\nu$ can, roughly,
be interpreted as discretized spatial coordinates. (Neither the
discretization in terms of $N$ sites nor the specific interpretation
of the indices will play any role throughout.)

Within the RMT description, space-type matrix elements will be
represented as $L^2A(\x,\x') \to NA_{\mu\nu}$ and the integration over
the coordinates of the upper system becomes a matrix trace,
$L^{-2}\int d^2x \to N^{-1}\sum_\mu$.  On the other hand, the dynamics
in the 2DEG spectrometer underneath is integrable-ballistic implying
that, as before, it has to be described in terms of the microscopic
Hamiltonian of the two-dimensional electron gas.  This type of hybrid
modeling, involving RMT in combination with a microscopic
Hamiltonian, does not pose any conceptual problems. In our basic
formulas, (\ref{current_uni}), (\ref{E24a}), and (\ref{central}), we
describe quantities assigned to the upper system through their RMT
representations whereas the correlation functions of the lower system
are given by the ballistic expression (\ref{E30}). Specifically, the RMT
representation of the correlation function $F^{[D]}$ of the upper
systems is given by
\begin{eqnarray*}
  F_1^{[D]}(\x,\x') \to F_{1\mu \nu}^{[D]}\equiv \langle
  G_{1\mu\nu}^{-}G_{1\nu\mu}^{+}\rangle,
\end{eqnarray*}
where $G_{\mu\nu}^\pm = (\epsilon\pm i0 - H \mp i
\Phi/2)^{-1}_{\mu\nu}$ are the RMT Green functions, and we have
assumed that the mean value of the perpendicular field
$B_\perp^+$ has been absorbed into the Hamiltonian $H$. The other
two correlation functions $F^{[d;C]}$ are given by similar
expressions with exchanged coordinates. Notice that none of the
functions $F$ actually depends on the coordinate arguments -- the
ergodic zero-dimensional nature of the dot -- which is why no
conflicts arise from the simultaneous appearance of RMT and truly
space dependent correlation functions in Eq.~(\ref{E24a}).

To be specific, let us  assume -- a condition easily met
experimentally -- that the electron dynamics on scales $L$ is
ballistic. The average tunneling conductance between the dot and
the extended system is again given by Eq.~(\ref{E7}). The only
difference is that, for a clean quantum dot with irregular boundary
scattering, the scale $\Gamma$ is given by $\Gamma^{\rm (b)} \sim v_{\rm
F}/L$ [where the superscript `(b)' stands for ballistic].

Turning to the fluctuations of the conductance, we first note that
for a GUE Hamiltonian, the cooperonic correlation function
$F^{[C]}=0$, and we are left with $F^{[d;D]}_{1\mu\nu} \equiv
F^{[d;D]}$, where the second denotation emphasizes the coordinate
independence. Substituting these expressions into
Eq.~(\ref{E24a}), temporarily setting the {\em
  parallel} magnetic field to zero, and integrating over coordinates
we obtain
\begin{eqnarray}
\label{C_RMT}
&&C_G(V,\Delta B) = 2\left(\frac{\Gamma^{\rm (b)}} {\epsilon_{\rm
      F}}\right)^2\left (\Re[F^{[d]}]+ \frac{\delta}{2\pi\Gamma^{\rm (b)}}\Re[F^{[D]}]\right)
      \,.\nonumber\\
\end{eqnarray}
The global suppression factor $(\Gamma^{\rm (b)}/\epsilon_{\rm F})^2\sim(k_{\rm
    F}L)^{-2}$ stems from the fact that an averaging over the area of the QD is
    intrinsic to the setup. The further suppression factor $\delta/\Gamma^{\rm (b)}\sim(k_{\rm
    F}L)^{-1}$ multiplying the diffuson contribution results
from the integration over the Green functions of the lower 2DEG.
Unlike the density-density contribution, $F^{[D]}$ is weighted by
Green functions taken at different coordinates. Integration over
these arguments leads to the $k_{\rm F}$-dependent suppression.

According to Eq.~(\ref{C_RMT}), the fluctuations of the tunneling
conductance are linearly related to the sum of two ergodic correlation
functions $F^{[d]}$ and -- multiplied by a small constant --
$F^{[D]}$. We next turn to the discussion of these objects. RMT
parametric correlation functions have been calculated within the
framework of the nonlinear sigma model~\cite{paracorr,TaSi95}. For
completeness we have included a brief review of these analyses in the
Appendix. The result, an integral representation of
the two functions $F^{[d;D]}$, is given by
\begin{eqnarray}
F^{[d]}(s;b) &=& \frac{1}{2b} \int\limits_0^\infty \frac{d y}{y} \, e^{is^+y} \,
  \big[e^{-by|y-2|}-e^{-by(y+2)}\big] \,, \label{A10}\\
F^{[D]}(s;b) &=& \frac{1}{2b^2} \int\limits_0^\infty \frac{d y}{y^3} \,
  e^{is^+y} \, \big[(by|y-2|+1)\,e^{-by|y-2|} \nonumber\\
&& \qquad\qquad -(by(y+2)+1)\,e^{-by(y+2)}\big]
\label{A11}\,.
\end{eqnarray}
In these equations, the correlation functions are characterized
through two dimensionless parameters, $s$ and $b$. The parameter
$$
s^+=\pi (V-V'+i0)/\delta_\uparrow
$$
measures the energetic mismatch between the two Green functions. (Throughout this section we will add to parameters of the upper/lower
system a superscript $\uparrow/\downarrow$, if necessary.)
The other parameter, $b\sim (N\delta_\uparrow)^{-2}{\,\rm tr \,}(\Phi^2)$,
describes the field difference. As shown in Ref.~\cite{paracorr}, this
parameter can unambiguously be related to the magnetic field threading
the dot.  Comparison of the RMT $\sigma$-model and its microscopic
counterpart leads to the identification,
$$
2b=\pi^2 C \phi^2 \,,
$$
where $C=\delta_\uparrow^{-2}\langle
(\frac{\partial\epsilon_n}{\partial\phi})^2\rangle$ describes the
sensitivity of levels to the applied field, and $\phi=\Delta B_\perp
L^2$ the magnetic flux threading the dot. For a disordered system, $C$
is proportional to the conductance $g$.

For general parameter configurations $(s,b)$, the integrals in
Eqs.~(\ref{A10}) and (\ref{A11}) cannot be done in closed form. In a number of relevant
limiting cases, however, simplifications are possible.  We first note
that for zero magnetic field difference, $\Delta B_\perp=0$, the
density-density correlator $F^{[d]}(s,b=0)$ is related to the
two-level correlation function (see, e.g., Ref.~\cite{Guhr98})
$$
R_2^u(s) = - \left({\sin s\over s}\right)^2
$$
through $\Re F^{[d]} =\pi\delta(s)-R_2^u(s)$ (the index `$u$' standing
for unitary). Thus, the density-density contribution to the {\em
  current} correlation function $C_I$ is simply proportional to the
level number variance $\Sigma_2^u(V/\delta_\uparrow)$. Similarly, the
real part of the zero-field diffuson correlation function is given by
$\Re F^{[D]}(s,b=0)=\pi\delta(s)$ (where the $\delta$-type dependence
on $s$ follows from the condition of particle number
conservation)\cite{fn_orth}. Notice that the $s=0$ singularity
displayed by the two correlation functions is of no relevance for our
theory. The reason is that for voltages $\Delta V \sim \delta_\uparrow
s < \Gamma_{\cal T}$ the tunneling approximation employed in
calculating the current is no longer applicable.  What happens for
smaller voltage differences cannot be figured out within the framework
of the present formalism. In principle, the $\sigma$-model description
can consistently be extended to include the effect of
{\em coherent} multiple tunneling. (A treatment neglecting the
coherence of higher-order tunneling processes would simply lead to a
complex energy shift, $s\to s+i\gamma$, where $\gamma = \pi
\Gamma_{\cal T}/\delta_\uparrow$ and, very much like dephasing
to a smearing of the singular low-energy profile of
the correlation functions.) The resulting theory, however, would be
significantly more complicated than the present formalism and we will
not discuss it any further (thereby paying the price that the very low
voltage regime remains out of reach).

We next focus on the dependence of the correlation functions on a
magnetic field. There are two different cases to be
distinguished: For small field strength, $b\ll1$, the effect of the
magnetic field is essentially nonperturbative (in the sense that the
correlation functions are nonanalytic in the inverse field strength
$\sim b^{-1}$). In this regime, the conductance correlation function
has to be computed through numerical integration of Eqs.~(\ref{A10})
and (\ref{A11}). The leading order term for small $b$ is linear in $b$ (i.e., quadratic in the field difference $\Delta B_\perp$).
For larger fields, $b\gg1$, a perturbative expansion in powers of
$b^{-1}$ and $s^{-1}$ leads to the results
$$
F^{[d]} =  \frac1{(2b-is^+)^2}, \qquad F^{[D]} = \frac2{2b-is^+}
$$
which would otherwise obtain from a perturbative diagrammatic
analysis (similar to the one outlined in the previous section).

We finally ask how the two contributions $F^{[d]}$ and $F^{[D]}$
contributing to the conductance correlation functions can be
distinguished.  In general, the factor $(k_{\rm F}L)^{-1}\ll1$ leads
to a massive suppression of the diffuson contribution as compared to
the density-density contribution. Since corrections of ${\cal
  O}((k_{\rm F}L)^{-1})$ have been neglected in applying random matrix
theory, anyway, an additional parameter is necessary to resolve the
diffuson contribution. As discussed previously within the context of
two extended systems, this is exactly what an {\em in-plane} magnetic
field does: Coupling only to the $D$-contribution, an in-plane field
difference $\B_\parallel^-=\B_\parallel - \B_\parallel'$ can be used
to selectively identify the $F^{[D]}$ correlation function.  In fact,
for finite $\B_\parallel^-$, the tunneling matrix elements pick up a
phase factor that modifies the subsequent integration over the Green
functions of the 2DEG. For small fields, this leads to
\begin{eqnarray*}
&& C_G(V,\Delta B_\perp;\Delta\q)-C_G(V,\Delta B_\perp;0)
\sim
\nonumber\\
&\sim& -\frac{c_g}{(k_{\rm F}L)^3}(|\Delta\q|L)^2 \,
F^{[D]}(s,b)+{\cal O}\LK(|\Delta\q|L)^4\RK \,,
\end{eqnarray*}
where $c_g$ is a constant of order unity.

The relevant field scale is one flux quantum through the area spanned
by the linear size of the QD, $L$, and the distance between the layers,
$d$, i.e.  $B_\parallel^{\rm c}\sim1/(dL)$. This field scale is
typically much larger than the characteristic scale of the
perpendicular magnetic field $B_\perp^{\rm c}\sim1/(\sqrt{g}L^2)$.

Before leaving this section, let us briefly comment on the connection
between the present analysis and previous experimental work by Sivan
et al.~\cite{Sivan94}. As mentioned in the introduction,
Ref.~\cite{Sivan94} investigated a setup dot/dot, where the second dot
was very small, with extreme level quantization. (That is,
in Ref.~\cite{Sivan94}, the role of our 2DEG is assumed by a single
quantized level of an ultrasmall device.) Arguing
semiquantitatively, Sivan et al.~related the statistics of the
tunneling conductance to density-density-type parametric correlation
functions.  Indeed, the experimental data turned out to be in good
accord with the RMT prediction, Eq.~(\ref{A10}), discussed above.
There are two differences to the presently discussed setup: first, the fact that in
our system a 2DEG is used as a spectrometer implies that the second,
diffuson-type correlation function plays a more important role than
in Ref.~\cite{Sivan94}. (In fact, this type of correlation function should
contribute to the data of Ref.~\cite{Sivan94}, too. Due to the fact, that
the current flows into a single level, however, this contribution is
minute and can safely be neglected.) Second, one may hope that the
spatial averaging involved in our formalism leads to a far-reaching
elimination of all non-spectral structures (as opposed to single-level
spectroscopy, where non-universal wave function characteristics may
affect the result). The price one has to pay is the suppression factor
$(k_{\rm F}L)^{-2}$ that is not present in the dot/dot setup.

Summarizing, we have found that the statistics of the
dot/2DEG-conductance (in a regime of broken time reversal invariance)
can be described in terms of two basic correlation functions
$F^{[d]}$ and $F^{[D]}$. As compared to the previously discussed case
of two extended systems, the information contained in $F^{[d;D]}$ is
now purely spectral. (All spatial structures have equilibrated due to
the ergodicity of the system.)  Indeed, these two functions are fully
universal in the sense that they depend only on the two basic
parameters $s$ and $b$ measuring bias voltage and perpendicular
magnetic field strength, respectively. There is one {\em non}-universal
element affecting the conductance fluctuations, viz., a geometry-dependent factor suppressing the contribution of $F^{[D]}$. Still the
contributions of the $d$- and $D$-correlation function can be
disentangled, namely, by measuring the dependence of the conductance
on a {\em parallel} magnetic field.  These results have been obtained
within a noninteracting theory. We next ask what might change once
Coulomb interactions are switched on.

\subsection{Coulomb interaction effects}\label{QD-C}

As far as quantum dots are concerned, the most important
manifestation of interaction phenomena is the Coulomb blockade:
due to the repulsive interaction,  it costs
extra energy $E_c$ to add an electron to the dot.
For an isolated dot, this charging energy is determined through the
self-capacitance $C$ through $E_c= e^2/(2C)$. However, for a
dot in close vicinity to an extended conducting system, a 2DEG, say,
an excess charge on the dot will be compensated for by
the accumulation of positive background charge in the large
system. Under such circumstances it is the geometric capacitance
{\em between} the systems that determines the charging energy. This is
the situation given presently.

For a planar dot/2DEG setup, the geometric capacitance estimates to
$C=\epsilon\epsilon_0L^2/d$ where
$\epsilon_0\approx8.9\cdot10^{-12}$ F$/$m and $\epsilon$ is the
dielectric constant of the filling medium. For GaAs,
$\epsilon\approx10$.  The capacitance determines the charging energy
$E_c$ that has to be compared with the other
characteristic energy scales of the problem. The smallest scale we
might hope to resolve is the single-particle-level spacing
$\delta_\uparrow$ of the dot. (It is this energy scale on which
non-perturbative structures in the parametric correlation functions
are observed.) Notice that both, $\delta_\uparrow = (\nu L^2)^{-1}$
and $E_c$, scale inversely with the dot area. Thus, it must be the
spacing between dot and 2DEG that determines the crossover criterion.
Specifically, with $\nu\approx3\cdot10^{10}$ ${\rm meV}^{-1}{\rm
  cm}^{-2}$, we find that $E_c \approx \delta_\uparrow$ for $d\approx
4$ nm.  Realistically, $d$ is somewhat larger than that, i.e., of the order of up to a few tens of nanometers. We thus conclude that
interaction effects are of relevance once one gets interested in low-energy structures of the order of the level spacing.

In previous sections, we have described fluctuations of tunneling
transport coefficients through disorder correlation functions ${\cal
  F}_0(\epsilon_1-\epsilon_2)\equiv\langle
A_0(\epsilon_1)A_0(\epsilon_2)\rangle$, where $A_0(\epsilon)$ are the
energy-dependent spectral functions and the subscript `$0$' means `noninteracting'. (In fact, we have largely focused
on the correlator of Green functions $F_0\sim \langle G_0^-G_0^+ \rangle$.
Presently, however, it will be more convenient to concentrate on the
spectral functions themselves. The two quantities $F_0$ and ${\cal F}_0$
are related through ${\cal F}_0 = \langle A_0A_0\rangle_c=2\Re\langle
G_0^-G_0^+\rangle_c= 2\Re F_0$.) As mentioned above,
interactions will mainly manifest themselves  through
global charging mechanisms. This sector of the Coulomb interaction does
not couple to the coordinate dependence of the correlation functions.
To simplify the notation, we therefore
temporarily suppress spatial coordinates in the notation.

Our main goal will be to show that interaction and disorder are
largely separable in the analysis of correlation functions. Yet,
unlike in the noninteracting case, it will no longer be sufficient to
compute the zero-temperature correlation functions and to account for
finite temperature effects in the end, through an integration over the
Fermi distribution functions. Instead, we have to work in a finite-temperature formalism from the outset.

To model the interaction we employ the `orthodox model', that is we
add a charging term
\begin{eqnarray}
H_c=\frac{e^2}{2C}(\hat N_{QD}-\bar N)^2\,,\label{Cou1}
\end{eqnarray}
to the Hamiltonian of the dot. Here $\bar N$ is the preferred
occupation number that can be set by the gate voltage.

To incorporate this term into the model, it is convenient to switch to
a functional integral formulation. Within this approach, it is
straightforward to see that the theory essentially splits into two
sectors: one that describes noninteracting Green functions (albeit
subject to some imaginary time dependent voltage $\sigma$), and an
interaction- and temperature-dependent weight function that controls
the fluctuations of $\sigma$. This approach of describing charging was
introduced by Kamenev and
Gefen~\cite{KaGe96}. In
the following, we shall briefly review its main elements, and apply it
to our present problem.

Within a fermion field integral approach, the imaginary time action
describing the quantum dot is given by
$$
S=\int d\tau\bar \psi(\partial_\tau+H_0-\mu)\psi
+E_c\int d\tau(\bar \psi\psi-\bar N)^2,
$$
where $\psi$ is a time- and position-dependent Grassmann field and
the rest of the notation is self explanatory. (Unless stated
otherwise, the notation $\bar \psi \psi \equiv \int d^2r \bar
\psi(\r)\psi(\r)$ contains a spatial integration.)  Decoupling, the
interaction by a Hubbard-Stratonovich transformation one obtains
the effective action
\begin{eqnarray*}
S[\psi,\sigma] &=& S[\sigma]+S_{\rm d}[\psi,\sigma],\\
&&S[\sigma]= \int d\tau  (\frac1{4E_c} \sigma^2 - i\bar N \sigma),\\
&&S_{\rm d}[\psi,\sigma]= \int d\tau
\bar \psi (\tau)\left(\partial_\tau + H -\mu + i
  \sigma\right)\psi (\tau),
\end{eqnarray*}
where $\sigma(\tau)$ is a scalar bosonic field.
Next, all but the static component $\sigma_0 \equiv \beta \int d\tau
\,\sigma$ are removed from the fermionic action, $S_{\rm d}$, through the gauge
transformation $\psi(\tau) \to \exp[-\int^\tau d\tau'
(\sigma(\tau')-\sigma_0)]\psi(\tau)$.  [That the static component
cannot be removed has to do with the fact that gauging out $\sigma_0$
would, in general, lead to a violation of the time-antiperiodic
boundary condition $\psi(\tau) = - \psi(\tau + \beta)$.] This makes
the action of the system insensitive to the time-dependent components of
the Coulomb field. However, the (imaginary time) Green functions
${\cal G}(\tau)$, we actually wish to compute, being non-gauge
invariant objects, pick up a gauge factor ${\cal G}(\tau,\sigma) \to {\cal
  G}(\tau,\sigma_0) B(\tau)$ to be specified momentarily. For temperatures
larger than the level spacing, the integration over the
static component $\sigma_0$ can be done in a saddle point
approximation. As a result, ${\cal G}(\tau,\sigma_0) \to {\cal
  G}_0(\tau,\bar \mu)\equiv {\cal G}_0(\tau)$, where $\bar \mu$ has the significance of an
effective (real) chemical potential determined through the optimal
occupation number $\bar N$.

We are thus led to consider the combination ${\cal G}(\tau) = {\cal
  G}_0(\tau) B(\tau)$. Roughly speaking, the physics of interactions
resides in the factor $B(\tau)$. Disorder, the external fields, etc.,
are contained in ${\cal G}_0={\cal G}_0(\x,\x';\tau)$. This is the
`splitting' of the zero-mode interaction theory into two components
mentioned above.

Transformation of ${\cal G}$ back to frequency space
obtains~\cite{KaGe96}
\begin{eqnarray}
{\cal G}(i\epsilon_n) &=&
-\frac{1}{2}\int\frac{d\omega'}{2\pi}\frac{d\epsilon'}{2\pi}B(\omega')
A_0(\epsilon')\times \nonumber \\
&&\times \frac{\coth\frac{\omega'}{2T}+\tanh\frac{\epsilon'}{2T}}{i\epsilon_n-\omega'-\epsilon'} \label{Cou2}\,,
\end{eqnarray}
where $A_0$ is the
spectral function associated with the (real-time) Green function $G_0$
and
$$
B(\omega)=2\sqrt{\frac{\pi}{TE_c}}\exp[-\frac{1}{4TE_c}(E_c^2+\omega^2)]\sinh\frac{\omega}{2T}.
$$
This representation holds for any particular realization of the
disorder and the external fields.

We next turn to the discussion of the correlator of two spectral
functions ${\cal F}(\omega;\epsilon)\equiv\langle
A(\epsilon+\frac{\omega}{2})A(\epsilon-\frac{\omega}{2})\rangle$.
Making use of Eq.~(\ref{Cou2}) and noticing that the disorder average
couples only to the functions $A_0$, we obtain
\begin{eqnarray}
 {\cal F}(\omega;\epsilon) &=& \frac{\pi}{TE_c}
\,e^{-\frac{E_c}{2T}}\int\frac{d\eta}{2\pi}\,{\cal
  F}_0(\omega-\eta)\,e^{-\frac{\eta^2}{8TE_c}}\times
\nonumber\\
&& \times\int\frac{dW}{2\pi}e^{-\frac{W^2}{2TE_c}}\frac{\cosh\frac{\epsilon}{T}+\cosh\frac{\omega}{2T}}{\cosh\frac{\epsilon\!-\!W}{T}+\cosh\frac{\omega\!-\!\eta}{2T}}
\,. \label{res-int}
\end{eqnarray}
Notice that, unlike ${\cal F}$, the non-interacting ${\cal F}_0$
depends only on the energy difference $\omega$.

Equation (\ref{res-int}) represents the most general form of our result for
the correlation function in the presence of a charging interaction. To
understand the structure of this expression, and to identify
physically distinct regimes, we realize that ${\cal F}$ depends
on four characteristic energy scales: the temperature $T$, the charging
energy $E_c$, the bias voltage $V$, and an intrinsic scale
$\epsilon_0$ over which the non-interacting correlator ${\cal F}_0$
varies. Notice that the dependence on $V$  is implicit, through the
limits imposed on the variables $\epsilon$ and $\omega$.

To exemplify the dependence of the correlator on these scales, let us
assume that ${\cal F}_0(\omega)$ is proportional to a delta function smeared over the
intrinsic level broadening $\Gamma_{\cal T}$, i.e., ${\cal F}_0(\omega) \sim \delta_{\Gamma_{\cal
    T}}(\omega)$ (as is the case, e.g.,
for the diffuson-contribution to the ergodic correlation functions
discussed above). In this limit,
\begin{eqnarray}
{\cal F}^{(\delta)}(\omega;\epsilon) &\sim&
\frac{1}{2\sqrt{TE_c}} e^{-\frac{\omega^2}{8TE_c}}f(\epsilon) \,,\label{F-delta}
\end{eqnarray}
where $f$ is some function that decays on a scale set by $T$. This
expression displays a feature common to {\em all} correlation
functions ${\cal F}$ affected by charging: Formerly sharp energy
dependences are broadened to Gaussians of width $\sigma=2\sqrt{TE_c}$. In other words, energetically sharp features of ${\cal
  F}_0$ are washed out and a lower limit to what can be resolved in an
experiment is set.

To make further progress, we have to distinguish between two different
regimes: (a) weak interaction or high temperature, $E_c \ll T$
and (b) strong interaction or low temperature, $E_c \gg T$. We
begin by discussing the first case, (a).

For $E_c \ll T$, interaction corrections are small and an
expansion to first order in the parameter $E_c/T$ obtains
\begin{eqnarray*}
{\cal F}(\omega;\epsilon)
&=& \bar{\cal F}_0(\omega)\left(1+ {\cal O}(E_c/T)\right) \,,
\end{eqnarray*}
where $\bar {\cal F}_0$ stands for an energy average of the noninteracting correlator ${\cal
  F}_0$, over a scale $2\sqrt{T E_c}$ (as in Eq.~(\ref{F-delta})). This means that weak
interactions leave the result essentially unaltered, albeit lowering
its spectral resolution.

We next turn to the discussion of case (b), $E_c \gg T$. Given that the
principal setup of the theory favors low temperatures, this regime is
more relevant than (a). On the other hand, it also has to be kept in
mind that the applicability of the theory~\cite{KaGe96} is limited to
temperatures $T>\delta$. Thus, the structures discussed below apply to
a temperature regime $\delta < T \ll E_c$.

In the extreme limit $T\to 0$~\cite{strictly}, the interactions
produce a hard Coulomb gap, i.e. ${\cal F}={\cal F}_0
\,\theta(\epsilon-|\omega|/2-E_c)$.  At finite temperatures the
gap becomes softer but its essential features remain robust: For
small bias $V$, the correlation function is strongly suppressed
while at large bias there are only small changes.  The relevant
limits are   (b1) $V \ll T \sqrt{E_c/T}$ and (b2) $T \ll E_c \ll V$.

In the first case, when the applied voltage is small,
Eq.~(\ref{res-int}) simplifies to
\begin{eqnarray*}
{\cal F}(\omega;\epsilon) &\simeq& c\,\bar {\cal
  F}_0(0)e^{-\frac{E_c}{2T}}(\cosh\frac{\epsilon}{T}+\cosh\frac{\omega}{2T}),
\end{eqnarray*}
i.e., the correlator is exponentially suppressed. Here $c$ is a
factor depending algebraically on $T/E_c$.

In the opposite case, when the applied voltage is large, one obtains
\begin{eqnarray*}
{\cal F}(\omega;\epsilon) &\simeq& \bar{\cal F}_0(\omega)-
\nonumber\\
&&-e^{-\frac{\epsilon-2E_c}{T}}\big(\bar{\cal
F}_0(\omega-)\, e^{-\frac{\omega}{2T}}+\bar{\cal F}_0(\omega+)\,
e^{\frac{\omega}{2T}}\big) \,,
\end{eqnarray*}
where the dominant contribution $\bar{\cal F}_0(\omega)$ is the
noninteracting ${\cal F}_0$ smeared out over energies $2\sqrt{TE_c}$
as before and $\omega\pm=\omega\pm 2E_c$.

Summarizing, large charging energies, $E_c > T$, change the
noninteracting theory in two different ways. First, to avoid the
Coulomb blockade, large bias voltages $V>E_c$ have to be
applied. Second, even for those voltages, a lower limit $2\sqrt{T
  E_c}$ on the maximal energetic resolution of the theory is
imposed. For the reasons outlined above, we believe that, at least for
sufficiently weak magnetic fields and impure samples, the effect of
other interaction mechanisms is relatively minor. At any rate, since the
functional dependence of the Coulomb blockade corrections
follows from Eq.~(\ref{res-int}), the applicability of the theory can
be put to test.

\section{Conclusions}\label{sec-conc}

In this paper, we have introduced an approach to exploring the three
basic two-particle correlation functions of mesoscopic physics, the
generalized diffuson $F^{[D]}$, the cooperon $F^{[C]}$,
and the density-density correlator $F^{[d]}$. The basic idea
is to monitor the current flowing between two parallel 2DEGs as a
function of three qualitatively different control parameters -- a
parallel magnetic field, a perpendicular field, and a bias voltage.

As compared to standard transport measurement architectures, the most
important advantage of the approach is that electronic correlations
are detected without disturbing the `bulk' system through {\em local}
contacts. Instead the entire planar electron system acts as an
`extended contact' whose spatial structure is, nondisturbingly,
scanned by means of the two magnetic fields; {\em spectral} electronic
structures are resolved by measuring the bias voltage dependence of
the current. Importantly, the relevant information carried by the
tunneling current is solely contained in its fluctuations. For example, we have
shown that the Fourier transform of the current-current correlation
function with respect to the external field, directly obtains the two spatially
resolved transport functions $F^{[D;C]}$. In contrast, previous
analyses of magnetotunneling currents, both experimentally and
theoretically, focused on the average current profile that is
unrelated to any `mesoscopic' type of information.

To exemplify the usefulness of the approach, we have considered
two different applications: tunneling between two extended 2DEGs
and tunneling from a quantum dot geometry into a 2DEG. As for the
former, we have shown how, at least in principle, the exponents
characterizing localization/delocalization transitions can be
extracted from the current data. In contrast, for a quantum dot
with ergodic dynamics, the focus is on spectral rather than on
spatial structures. We have explicitly worked out the connection
between the tunneling current statistics and various parametric
correlation functions (a connection previously used on a
semiquantitative basis to interpret the data of the
experimentreported in Ref.~\cite{Sivan94}) and discussed how these correlations
can be monitored by changing external field and bias voltage.

As with any other tunneling setup, Coulomb interactions are likely to
change the outcome of the noninteracting theory. We have provided
evidence in favor of the Coulomb blockade being the most relevant
interaction mechanism.  The presence of the Coulomb blockade will
result in two principal effects: first, it forces one to use bias
voltages in excess of the blockade threshold.  Second, the spectral
structure of the correlation functions is washed out. This reduces the
information content that can be extracted from the tunneling current
statistics. The extent to which these obstructive mechanisms affect
the theory is set by the Coulomb charging energy. In the present
system, the latter is largely determined by the inverse of the
interlayer capacitance which, owing to the extended geometry of the
systems, is small. Thus, charging phenomena will be less pronounced
than in small islands. All in all, we believe that for realistic
values of the relevant system parameters, a significant parameter
range over which electronic correlations can be resolved through the
current approach remains.

{\sc Acknowledgments:}
We are grateful to B.L.~Altshuler and L.S.~Levitov for valuable discussions.
This work was supported by the Deutsche Forschungsgemeinschaft (DFG).
\begin{appendix}

\section{RMT Correlation Functions}\label{app-RMT-corr}

In this appendix, we briefly recapitulate
the essential steps of the the nonlinear $\sigma$ model (NL$\sigma$M) analysis of
RMT parametric correlation functions. The goal is to compute
correlation functions of an advanced and a retarded Green function
taken at different energies and magnetic fields. To model the field
difference, we assume that the Hamiltonian of the retarded/advanced Green
function is given by $H\pm i\Phi/2$.
Here $\Phi$ is some antisymmetric matrix
that can be modeled in a variety of different ways~\cite{crossover}.
We are thus led to consider correlation functions of the type
\begin{eqnarray*}
&&F_{\mu\nu\mu'\nu'}
\equiv \langle (\epsilon^++\frac\omega2-H-i\Phi/2)^{-1}_{\mu \nu} \\
&&\hspace{3.0cm} (\epsilon^--\frac\omega2-H+i\Phi/2)^{-1}_{\mu'\nu'}\rangle.
\end{eqnarray*}
Correlation functions of this type can conveniently be calculated from
the functional integral~\cite{NLSM}
\begin{eqnarray}
&&\int{\cal D}[\psi,\bar \psi]\, e^{-S[\psi,\bar \psi]}(\dots)\,,\label{Z_func}\\
&&\qquad\qquad S = i\bar \psi\Lambda^{1/2}\{\epsilon-H - ({\omega^+\over 2}-i\Phi)\Lambda\}\Lambda^{1/2}\psi\,.\nonumber
\end{eqnarray}
Here $\Lambda=\sigma_3^{\sc ar}$ is a Pauli matrix in the
`advanced-retarded' space, $\omega^+=\omega+i0$, and the ellipses
stand for pre-exponential terms specific to the index configuration of
the correlation function under consideration~\cite{NLSM,crossover}.
Following a by now standard procedure~\cite{NLSM,crossover,paracorr},
the average over the RMT Hamiltonian then leads to
the effective action of the NL$\sigma$M,
\begin{eqnarray}
 S_{\rm eff}=
\frac{s^+}{2}\,{\rm STr}( Q\Lambda )- \frac{b}{4}\,{\rm STr}(
Q\Lambda)^2\,,
\label{S_eff}
\end{eqnarray}
where $Q$ is a four-dimensional supermatrix subject to the constraint
$Q^2=1$. The parameters, $s$ and $b$, appearing in this expression
measure the `mismatch' of our two Green functions. Their relation to system parameters is discussed in Sec.~\ref{QD-noC}.

At this stage the computation of the parametric correlation functions
$F^{[d;D;C]}$ amounts to an integral over the four-dimensional matrix
$Q$. The explicit calculation~\cite{paracorr,TaSi95} leads to
$F^{[C]}=0$ and the two one-dimensional integral representations
displayed in Eqs.~(\ref{A10}) and (\ref{A11}) in the main text.

\end{appendix}

\end{multicols}

\begin{references}
\bibitem{fn_tunneling} Here, `weak' means that the typical energy scale 
  associated 
  with the tunneling, $\Gamma_{\cal T}\propto|{\cal T}|^2$, is much
  smaller than the other relevant energy scales. In principle, the effect of 
  higher order
  tunneling processes can be systematically taken into account. Within an
  approximation whereby higher-order tunneling is treated as an
  incoherent process this leads to the level smearing mentioned in
  the text (see also Sec.\protect~\ref{sec-spectral} below.) Due
  to the assumption, $V > \Gamma_{\cal T}$, a more complicated, coherent
  description is not necessary.
\bibitem{fn-dis} Correlations in
the disorder potentials might become important if the impurities were
placed between the 2DEGs which, in a usual setup, is not the case. In
principle, long-ranged fluctuations of the Coulomb potential due to
remote impurity sites may be felt by both layers simultaneously. In
practice, however, these potentials are strongly screened by the
layers themselves which implies that no significant interlayer
correlation remains. Notice that correlation effects become
`significant' once the amplitude of the correlated part of the
scattering potential parametrically exceeds that of the uncorrelated
potential.
\bibitem{fn-drag} Typically, the relevant time scale
 $\tau_{1\leftrightarrow 2}$ for Coulomb drag is much larger than the momentum
relaxation $\tau$ within each layer, see, e.g., Ref.~\cite{ZMcD93b}.
\bibitem{fn_orth} It is straightforward to generalize the
               $\Delta B_\perp=0$ result to the orthogonal case by
               replacing $R_2^u(s)$ in $F^{[d]}$ by the corresponding
               function $R_2^o(s)$ and multiplying the diffuson
               contribution by a factor of 2 ($F^{[C]}=F^{[D]}$ at
               $B_\perp=0$).
\bibitem{strictly} Strictly speaking, the limit $T\to 0$ is not compatible 
with the condition $T\gg\delta$, but it yields
                    qualitatively correct results.
\end{references}
\end{document}